\documentclass[%
reprint,
superscriptaddress,
frontmatterverbose,
 amsmath,amssymb,
prl,
]{revtex4-1}

\usepackage{amsfonts,amsmath,amssymb,amsthm}

\usepackage{soul,xcolor}
\usepackage{graphicx}
\usepackage{dcolumn}
\usepackage{bm}

\usepackage{hyperref}
\hypersetup{
	pdfborder={0 0 0}
	pdfnewwindow=true,      
	colorlinks=true,       
	linkcolor=blue,          
	citecolor=blue,        
	filecolor=blue,      
	urlcolor=blue,          
}
\usepackage{color}
\usepackage{amsmath}

\usepackage{lineno}
\makeatother
\usepackage{ulem}

\newcommand{\be}{\begin{equation}}
\newcommand{\ee}{\end{equation}}

\usepackage{xr}
\newcommand\agf{AgF$_2$}
\setstcolor{red}

\begin{document}

\title{A silver(II) route to  unconventional superconductivity}

\author{Xiaoqiang Liu}
\thanks{Equal contribution}
\affiliation{International Center for Quantum Materials, School of Physics, Peking University, Beijing 100871, China}

\author{Shishir K. Pandey}
\thanks{Equal contribution}
\affiliation{International Center for Quantum Materials, School of Physics, Peking University, Beijing 100871, China}

\author{Ji Feng}\email{jfeng11@pku.edu.cn}
\affiliation{International Center for Quantum Materials, School of Physics, Peking University, Beijing 100871, China}
\affiliation{Collaborative Innovation Center of Quantum Matter, Beijing 100871,
China}
\affiliation{CAS Center for Excellence in Topological Quantum Computation, University of Chinese Academy of Sciences, Beijing 100190, China}

\begin{abstract}
The highly unusual divalent silver in silver difluoride (\agf) features a nearly square lattice of Ag$^{+2}$ bridged by fluorides. 
As a structural and electronic analogue of cuprates, its superconducting properties are yet to be examined. 
Our first principles electronic structure calculations reveal a striking resemblance between \agf~and the cuprates. 
Computed spin susceptibility shows a magnetic instability consistent with the experimentally observed antiferromagnetic transition. 
A linearized Eliashberg theory in fluctuation-exchange approximation shows an unconventional singlet $d$-wave superconducting pairing for bulk \agf~at an optimal electron doping. The pairing is found to strengthen with a decreasing interlayer coupling, highlighting the importance of quasi-2D nature of the crystal structure. These findings place \agf~in the category of unconventional high-$T_\text{C}$ superconductors, and its chemical uniqueness may help shed new lights on the high-$T_\text{C}$ phenomena.
\end{abstract}

\maketitle

\textsf{\textit{Introduction}}. 
Superconducting properties of high $T_\text{C}$ cuprates emerge from an intricate interplay between electronic, lattice and spin degrees of freedom \cite{cu1,cu2}.
The cuprate crystal structure is generally derived from the perovskite type structure, featuring a few universal themes. 
Structurally they all contain quasi two-dimensional (2D) CuO$_2$ sheets. 
Their normal state electronic structure near the Fermi energy is dominated by a single band derived from Cu-$d$ orbitals \cite{cu1,cu2,cu3,cu4,cu5,cu6}.
In sharp contrast to conventional superconductors based on electron-phonon coupling assisted Cooper pair formation, superconductivity in cuprates is believed to be driven largely by strong electronic interactions \cite{cu6}.
The quasi-2D nature of crystal structure limits  electronic modes in the out-of-plane direction, resulting in reduced screening and enhanced interaction that are essential to high-$T_\text{C}$ superconductivity. Understanding obtained from the extensive studies of structural, electronic and superconducting properties of cuprates has led to discoveries of new superconducting materials \cite{fe11,cu6,bg}.
Insights into the interplay of geometric and electronic structure are key to discovery of novel superconductors.
Clearly, it is then attractive to assay materials that resemble cuprates, both structurally and electronically, for potential novel superconductivity.

\begin{figure}[b]
\centering
\includegraphics[width=7.cm]{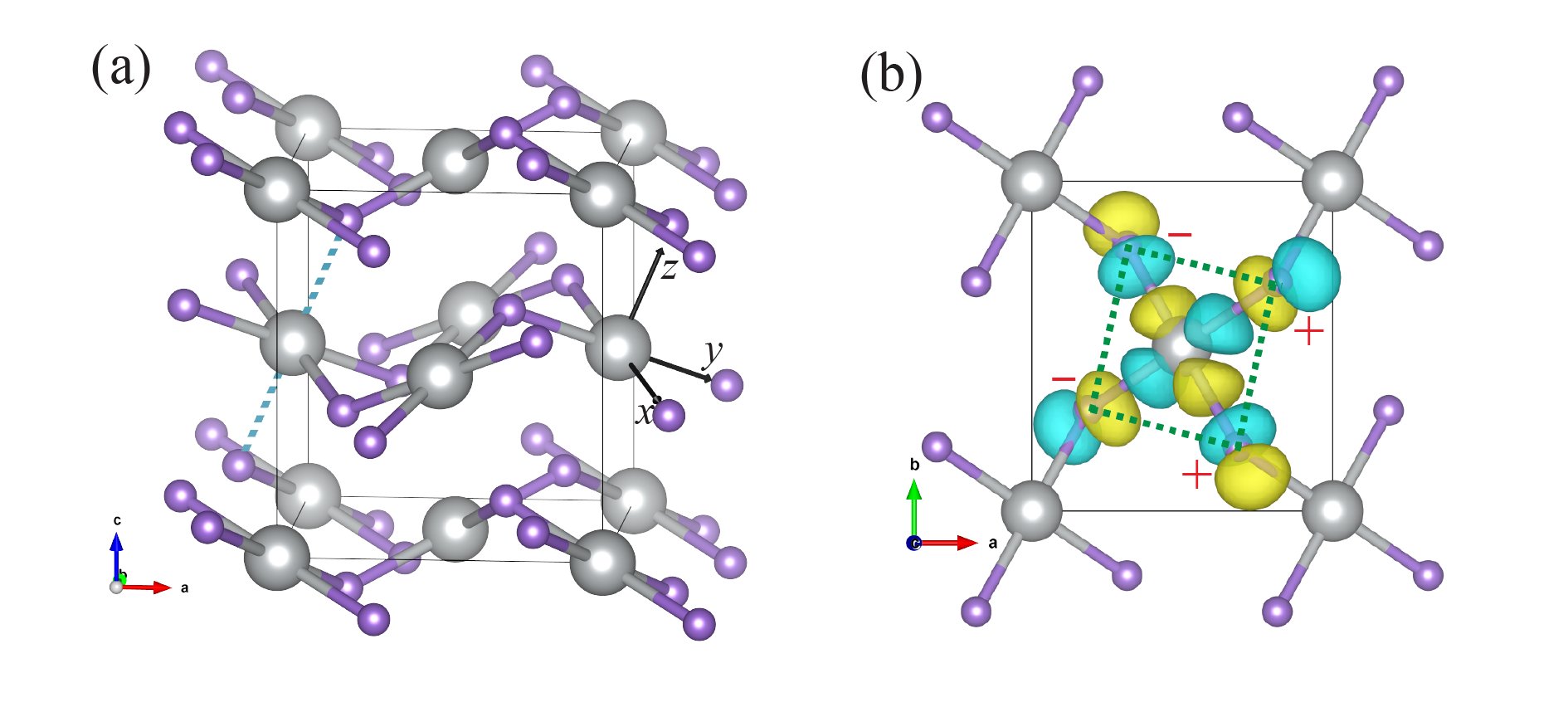}
\caption{(a)  AgF$_2$ crystal structure. Purple and gray balls are F and Ag, respectively. Blue dashed lines through one of the Ag(II) indicate out-of-plane Ag-F bonds in a AgF$_6$ octahedron. The black arrows labeled $x,y,z$ indicate the local coordinates used to describe $d$-orbitals on Ag. In (b), green dash-lined box highlights a AgF$_4$ unit. +/- indicate out-of-plane displacements of fluoride ions. The $d_{x^2-y^2}$ Wannier orbital on the central Ag is shown.}
\label{fig:1}
\end{figure}

Materials hosting divalent silver are extremely scarce in comparison with monovalent silver compounds.
Silver difluoride (\agf) has been synthesized from AgNO$_3$, anhydrous hydrogen fluoride treated with K$_2$NiF$_6$ and elemental fluorine, with silver ion Ag(II) in a highly unusual divalent state despite 
relatively large second ionization potential compares to the first one \cite{agf1,agf5}. More interestingly, \agf~resembles cuprates's parent phase La$_2$CuO$_4$ in its geometric, electronic and magnetic structures. An \agf~sheet of the bulk crystal is structurally similar to a CuO$_2$ sheet, with similar pattern of out-of-plane displacement of anion atoms as shown in Fig. \ref{fig:1}(a).
The divalent Ag features a 3$d^9$ valence shell, iso-valent to cuprates. 
The antiferromagnetic ground state of charge neutral \agf\, is a charge-transfer insulator,  which again is a familiar scenario in cuprates.

It is then a natural and tempting question whether \agf~will turn superconducting once metallized upon doping, like the cuprates. It is therefore the purpose of this work to study whether interaction can drive a superconducting transition in \agf, and what the ensuing pairing symmetry will be.
We start with an investigation of crystal and electronic structure of \agf~and its resemblance to the archetypal cuprate, the orthorhombic La$_2$CuO$_4$ \cite{cu6}.
A comparison of the crystal and electronic structures obtained from first principles calculations establishes a compelling structural and electronic resemblance between these compounds.
A multiband Hubbard model is constructed 
from which the spin susceptibility of \agf~within the random-phase approximation reveals an antiferromagnetic instability in accordance with experiments. 
Employing the fluctuation-exchange approximation and solving the linearized Eliashberg equations,  we obtain the superconducting pairing strength ($\lambda$) and symmetry. 
A phase diagram is obtained by calculating $\lambda$ at various carrier doping level and Hubbard $U$ values. 
The strongest superconducting pairing is obtained at 5\% electron doping for bulk \agf, with a dominating singlet $d_{xz}$ symmetry.
We find that the superconducting pairing strength is gradually noted to increase with decreasing interlayer coupling. We attribute this effect to the renormalization of electron-electron correlations with decreasing out-of-plane coupling.

\textsf{\textit{Electronic structure}}.
The structure of \agf~can be viewed as a stack of Ag-F square-planar networks resembling the cuprate planes in La$_2$CuO$_4$ \cite{cu6} as shown in Fig.~\ref{fig:1}(a). 
Similar to the low-temperature polymorph of La$_2$CuO$_4$ \cite{cu6}, \agf~has an orthorhombic crystal 
structure with each Ag(II) in a distorted octahedral crystal field of six nearest-neighbor F$^-$ ions \cite{agf2,agf3}. However, unlike in a perfect octahedral coordination, the out-of-plane Ag-F bonds are elongated by 24\% relative to the in-plane ones as shown by blue dashed lines on one of the Ag(II) in Fig.~\ref{fig:1}(a), leaving Ag(II) 4-coordination in a \agf~unit. This again resembles La$_2$CuO$_4$ in which there is a 27\% elongation of the out-of-plane Cu-O bonds. 
These four F$^-$-coordinated Ag(II) form AgF$_4$ unit within the square-planar network, as indicated by green dash-lined box in Fig.~\ref{fig:1}(b). A significant deviation of \agf~structure from La$_2$CuO$_4$ comes from the tilting of this AgF$_4$ unit by a large angle $\sim25^\mathrm{o}$, and hence the plane is puckered as shown in Fig.~\ref{fig:1}. 
This tilt of CuO$_4$ in La$_2$CuO$_4$ is much gentler ($\sim5^\mathrm{o}$). 
The TM-anion-TM (TM = Ag or Cu) angles in the square-planar structure are $\sim130^\mathrm{o}$ for \agf, which is $\sim173^\mathrm{o}$ in La$_2$CuO$_4$. 
This distortion from the ideal $180^\mathrm{o}$ angle is expected to manifest itself in the superexchange interaction, and therefore the temperature of magnetic ordering.
Indeed, the Ne\'el temperature (T$_N$) is 300 K for La$_2$CuO$_4$ and 163 K for \agf~\cite{cu7_dope,agf4}.
Given the striking similarities of structural and magnetic properties of \agf~with that of La$_2$CuO$_4$ and the subtle difference, investigation of its
electronic properties in context of superconductivity is warranted.

\begin{figure}[ht]
\centering
\includegraphics[width=8.cm]{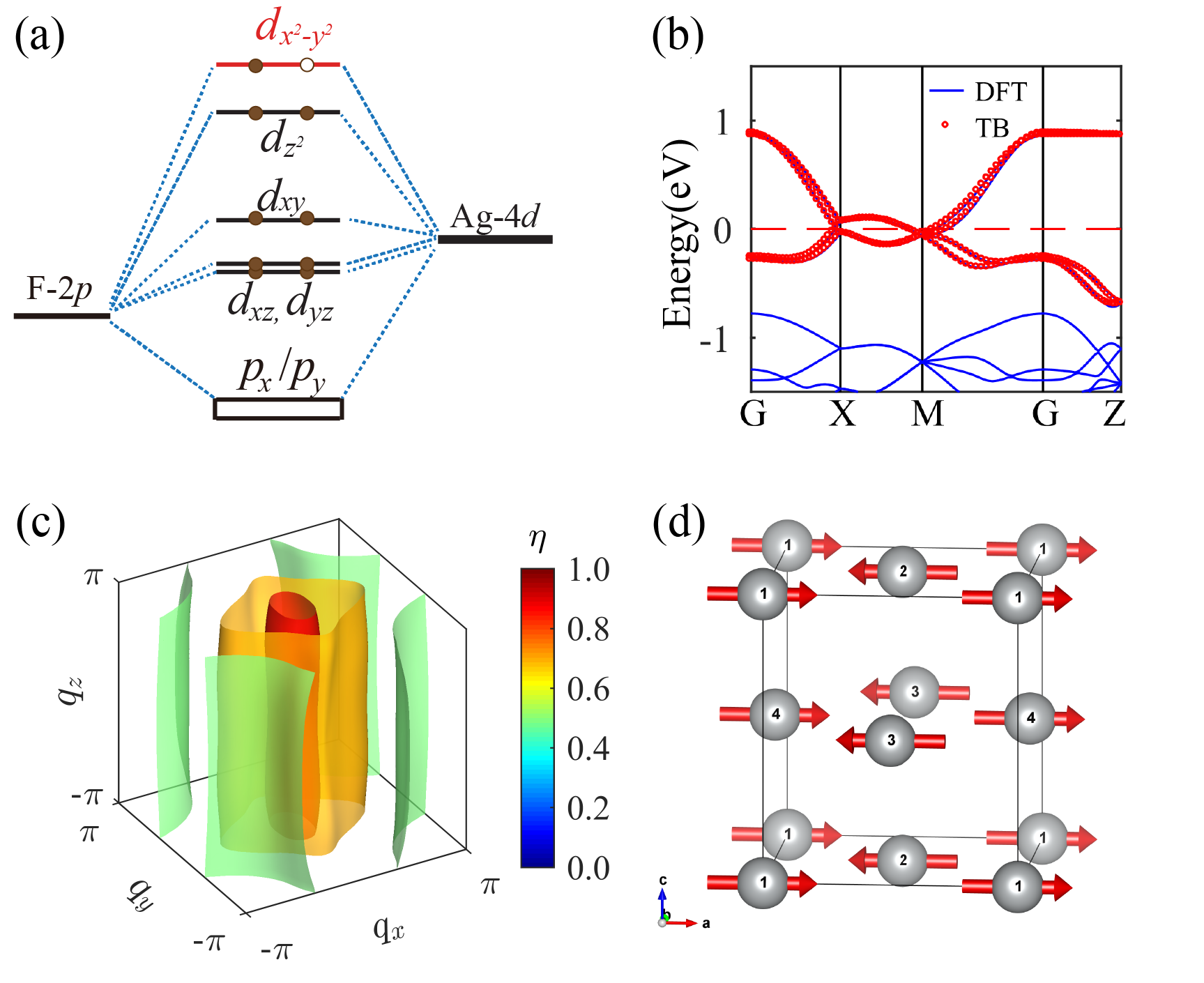}
\caption{
(a) Schematic energy level diagram of Ag(II) in the crystal field of F atoms.  
(b) Non-spin polarized band structures of \agf~from first principles calculation (DFT) and Wannier interpolation based tight binding model(TB), near the Fermi energy.
(c) $\eta(\bm{q})$ for undoped bulk \agf~with isovalue 0.5(cyan), 0.7(yellow) and 0.9(red). Here, $U$=0.44 eV and $T$=14 meV. 
(d) Magnetic structure of \agf~obtained from RPA. Red arrows represent the magnetic moment on Ag atoms. }
\label{fig:2}
\end{figure}

As discussed earlier, a single $d_{x^2-y^2}$ orbital for \agf~shown in Fig.~\ref{fig:1}(b) dominating the low energy space near the Fermi level is one of the most 
prominent characteristic feature similar to the cuprates. This is schematically shown in Fig.~\ref{fig:2}(a) where partially filled $d_{x^2-y^2}$ shown in red, contributes at the Fermi level.
In the octhedral crystal field, the $d$ orbitals are split into a triply degenerate $t_{2g}$ set and a doubly degenerate $e_g$ set. 
Deviation from perfect octahedral symmetry described earlier lifts the degeneracy of $e_g$ orbitals with $d_{z^2}$ being lower in energy than the in-plane $d_{x^2-y^2}$. 
The occupied anion 2$p$ orbitals are situated deep below the Fermi level.  
Non spin-polarized band structure of \agf, calculated using the density-functional theory, shown in Fig.~\ref{fig:2}(b) exhibits  features akin to cuprates. For calculational details, refer to the Supplemental Material(SM) \cite{sm}.
The low-energy excitations are dominated by the half-filled $d_{x^2-y^2}$ on Ag,  and are well separated 
from all other bands. Thus when constructing a tight-binding  model, it is justified to include simply one $d_{x^2-y^2}$-like Wannier orbital per Ag.
Once the Coulomb interaction is included \agf~becomes charge-transfer antiferromagnetic insulator similar to the cuprates in a scenario discussed also in 
Jakub et al \cite{agf1}.

\textsf{\textit{Interaction-mediated superconductivity}}.
To investigate the effect of interaction on the magnetic order and potential superconductivity, we construct a multiband Hubbard model,
\begin{equation}
\label{eqn:eq1}
H=H_0+H_U=\sum_{ijll'\sigma}t_{ij}^{ll'}c_{il\sigma}^{\dagger}c_{jl'\sigma}
+U\sum_{il}n_{il\uparrow}n_{il\downarrow},
\end{equation}
where $i, l$ and $\sigma$ are lattice, orbital and spin indices, respectively,  and $c$ and $n$ are Fermion annihilation and number operators, respectively.
The low-energy bands are described by tight-binding Hamiltonian $H_0$, in which hopping amplitudes are derived from the maximally-localized Wannier function approach \cite{wannier90} (see Table S1 in SM \cite{sm}).  The resultant band structure is shown in  Fig.~\ref{fig:2}(b), where the four $d_{x^2-y^2}$ bands (four Ag per unit cell) from the tight-binding model fit the first principles bands well. The intra-orbital Hubbard parameter $U$ is determined by estimating the Ne\'el temperature in a random-phase approximation (RPA), as described next.

Within the RPA \cite{rpa1,rpa2,rpa3,rpa4,rpa5}, the charge ($\chi^c$) and spin ($\chi^s$) susceptibilities  are given by,
\begin{equation}
\begin{array}{cc}
\chi^{c}(q)= & \left[1+\chi^{0}(q)U^{c}\right]^{-1}\chi^{0}(q)\\
\chi^{s}(q)= & \left[1-\chi^{0}(q)U^{s}\right]^{-1}\chi^{0}(q)
\end{array},
\end{equation}
where  $q=(\bm{q},\omega)$, $\chi^{0}$ is the bare susceptibility, and $U^{c}$ and $U^{s}$ are the interaction matrices in the charge and spin channels, respectively \cite{sm}.
The onset of spin instability is detected by the condition $\left|1-\chi_0(\bm{q},0)U^{\mathrm{s}}\right|$ = 0, which happens when  the maximum eigenvalue of $\chi_0(\bm{q},0)U^{\mathrm{s}}$ (denoted by $\eta(\bm{q})$)  becomes unity at any $\bm{q}$. 
The ensuing divergence of $\chi^s$ leads to a magnetic phase transition. 
The vector $\bm{q}^*$ and temperature $T_{\text N}$ at which $\eta(\bm{q^*})=1$ are the N\'eel temperature and propagation vector of the spin 
pattern, respectively. 
The spin pattern corresponding to a $\bm{q}^*$ is determined by diagonal elements of the eigenvector $\xi(\bm{q^*})$ corresponding to  $\eta(\bm{q^*})$. We use a mesh of $48\times48\times48$ for Brillouin zone sampling in all our calculations on the Hubbard model.

The Hubbard $U$ is estimated to be 0.44 eV, by matching the experimentally observed  N\'eel temperature($\sim163$ K) of \agf~(Fig. S1 in SM \cite{sm}).
Isosurfaces of $\eta(\bm{q})$ drawn in Fig.~\ref{fig:2}(a) for undoped bulk \agf~at $T=14$ meV ($\sim163$ K) show a strong anisotropy corresponding to a strong intra-layer and a weak interlayer magnetic exchange interactions.
Henceforth, we focus on $q_z=0$ plane in current analysis. 
 The maximum value of $\eta(\bm{q})$ is found to lie along the $q_z$-axis. Thus, for a weak interlayer coupling when restricted only in the $q_x$-$q_y$ plane, the maximum value of 
 the spin susceptibility is attained at $\bm q =0$. 
The computed eigenvectors $\xi_{ll} (\bm q =0)$ yield an antiferromagnetic order shown in Fig.~\ref{fig:2}(d), consistent with the experimentally established N\'eel state in \agf~\cite{agf4}.

When doped with a carrier concentration that readily suppresses the magnetism, a cuprate goes metallic exhibiting various 
kind of instability such as charge and spin fluctuations at low temperatures due to Fermi surface reconstruction.  The Hubbard models have been used extensively to explain the superconductivity in doped cuprates \cite{fe11}. Keeping the striking resemblance of \agf~with cuprates,  similar approach of metallization by carrier doping 
applies in \agf~as well. Thus, having a model capable of describing the  magnetic instability and order of \agf, we go on to a scrutiny for potential superconductivity mediated by spin fluctuation. 
The fluctuation-exchange approximation(FLEX) \cite{FLEX1,FLEX2} is employed to describe the effective electron-electron interaction $\Gamma(q)$ given by,
\begin{equation}
\Gamma(q)=\gamma U^{s}\chi^{s}(q)U^{s}-\frac{1}{2}U^{c}\chi^{c}(q)U^{c}+\frac{1}{2}(U^{s}+U^{c})
\end{equation}
with $\gamma=\frac{3}{2}$ for the singlet channel and $\gamma=-\frac{1}{2}$ for the triplet channel. An effective pairing between the electrons on the Fermi surface arising from spin and/or charge fluctuations can result in the formation of Cooper pairs. To describe the pairing instability of this type, the linearized Eliashberg equation is solved in the weak-coupling regime,
\begin{equation}
\label{eqn:eq2}
\lambda\phi_{mn}(\bm{k})=-\frac{1}{N}\sum_{\bm{k}'}\sum_{\mu\nu}\Gamma_{\mu\nu}^{mn}(\bm{k},\bm{k}')F_{\mu\nu}(\bm{k}')\phi_{\mu\nu}(\bm{k}')
\end{equation}
where $F_{\mu\nu}(\bm{k}')$ is a factor arising from summing the product of Green's functions over Matsubara frequencies and $m,n,\mu,\nu$ are band indices. 
$\phi_{mn}(\bm{k})$ is the order parameter  of superconducting phase \cite{sm}. Eq.(\ref{eqn:eq2}) then is an eigenvalue equation. 
The largest eigenvalue $\lambda_\mathrm{max}$ becomes unity at superconducting $T_\text{C}$ and can be used to gauge the relative pairing strength near the $T_\text{C}$.

\begin{figure}[ht]
\centering
\includegraphics[width=8.cm]{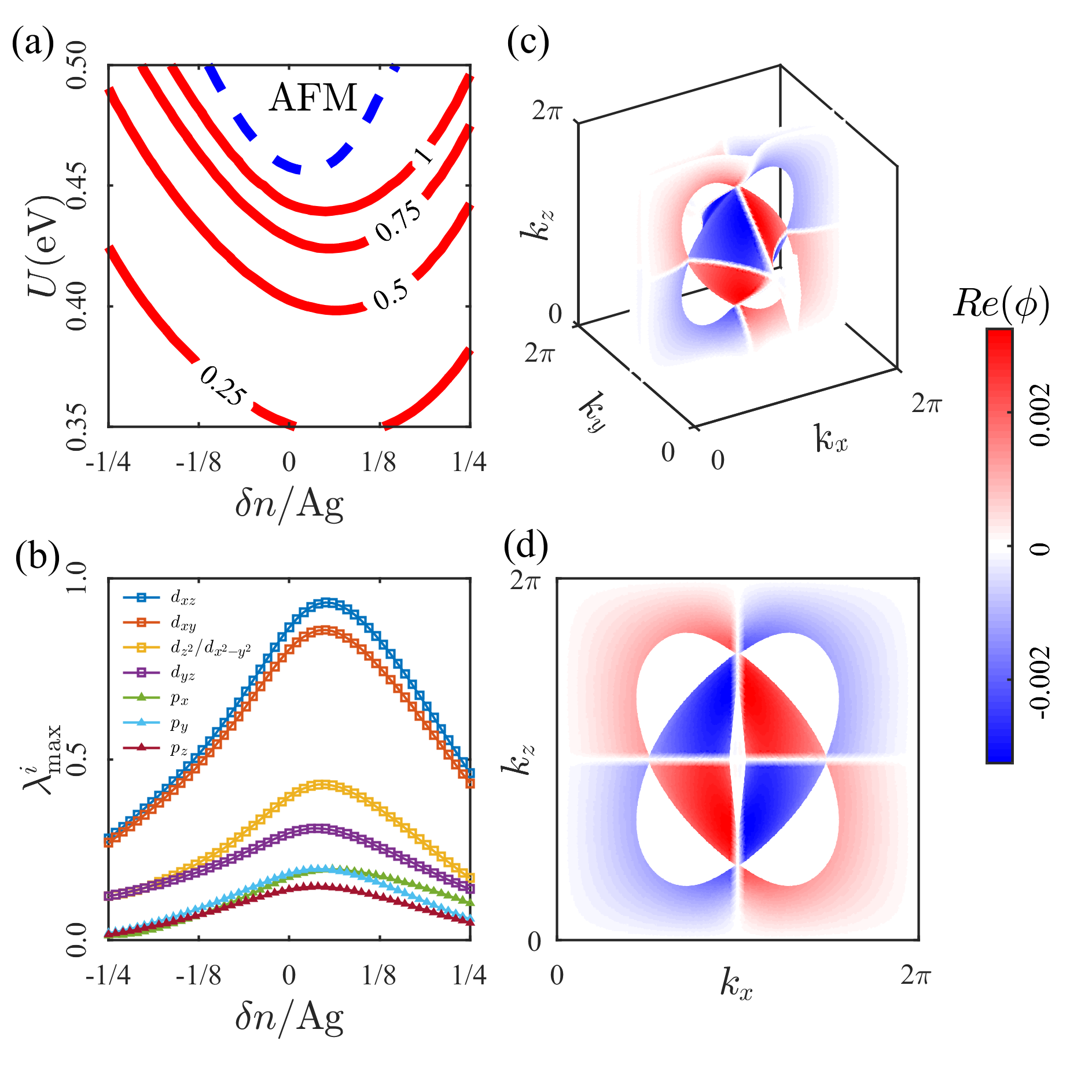}
\caption{(a) Contour plot of $\lambda_{\mathrm{max}}$ of the linearized Eliashberg equation at $T=30$ meV. Blue dash line correspond to $\eta(\bm{q})$ reach to unity at any $\bm{q}$. (b) Doping dependence of $\lambda^i_{\mathrm{max}}$ for several pairing symmetry of bulk \agf~ at 
$U=0.44$ eV and $T=30$ meV. Square representing the singlet and triangle representing the triplet channels.  (c) Three-dimensional and (d) projection on $k_x$-$k_z$  plane, of the order parameter  $\phi(\bm{k})$ at Fermi surface of one of the \agf~layer with $\delta n/$Ag$=0.05$ for bulk \agf.
}
\label{fig:3}
\end{figure}

Eq.(\ref{eqn:eq2}) is solved for various doping levels and  $U$ values at $T=30$ meV. 
Fig.~\ref{fig:3}(a) shows the contour plot for $\lambda_\mathrm{max}$ in the $U-\delta n$ parameter space. 
As observed in Fig.~\ref{fig:3}(a) the superconducting pairing strength increases with increasing $U$ at a given doping, underlining the importance of electronic correlation for potential  superconductivity in this compound. 
The symmetry of pairing can be identified by assigning each solved $\phi_{mn}(\bm{k})$ to an  
irreducible representation of the $D_{2h}$ point group, symmetry group of the bulk \agf. 
Corresponding to each irreducible representation $i$ (Table S2 in SM \cite{sm}), the largest eigenvalue is denoted by $\lambda^i_{\mathrm{max}}$.
Fig.~\ref{fig:3}(b) shows the doping dependence of $\lambda^i_{\mathrm{max}}$ for various pairing symmetry at $U=0.44$ eV.  
One can find that the singlet $d$-wave pairings have significantly higher strength than triplet $p$-wave pairings and the leading pairing symmetry is singlet $d_{xz}$-type wave 
throughout the $U-\delta n$ parameter space shown in Fig.~\ref{fig:3}(a).
Moreover, the hole doping readily decreases $\lambda$, while the electron doping  tends to increase $\lambda$ at first, reaching a peak value at an optimal doping of 5\% beyond which further doping tends to reduce $\lambda$. 

To circumvent the gauge problem \cite{sm} of $\phi_{mn}(\bm{k})$ for degenerate bands, we define $\phi(\bm{k})$ as 
\begin{equation}
\phi(\bm{k})=\sum_n\phi_{nn}(\bm{k})\delta(\varepsilon_{n\bm{k}}-\varepsilon_{\mathrm{F}})
\end{equation}
which describes the nature of order parameter on Fermi surface. Here $\varepsilon_{n\bm{k}}$ and $\varepsilon_{\mathrm{F}}$ are band energy and Fermi energy respectively. 
Energy cutoff of 5 meV is considered for evaluation of $\delta$ function in our calculations. 
We plot the real part of $\phi(\bm{k})$ corresponding to $\lambda_{\mathrm{max}}$ for one of the \agf~layers with $U=0.44$ eV and optimal doping of 5\%.
This is shown in Fig.~\ref{fig:3}(c) in the 3-dimensional Brillouin zone, and Fig.~\ref{fig:3}(d) shows a projection onto the $k_x$-$k_z$ plane. 
One encouters nodes crossing $k_x=\pi$ or $k_z=\pi$ planes indicating a $d_{xz}$-wave pairing. Hence, it can be concluded that the bulk \agf~crystal becomes unstable to a $d$-wave pairing induced by spin fluctuation.

\textsf{\textit{Interlayer coupling}}.
As discussed in the beginning, quasi-2D nature of the crystal structure of cuprates is one of the factors favoring its high 
$T_\text{C}$ \cite{cu_mono,mono}. 
In the case of \agf, although Ag-F layers resemble the copper oxide sheets, the separation of these planes is 2.91 \AA{}, much smaller than what is observed for La$_2$CuO$_4$ (6.6 \AA{}) as well as other cuprates. 
Consequently, the effect of interlayer coupling on superconducting properties of \agf~clearly warrants further study.
In other words, \agf~provides a good platform to investigate the role of quasi-2D nature of crystal structure in superconducting properties. 
Additionally, monolayer or few-layer samples more prone to doping  by techniques such as field or electrolytic gating \cite{gating}, 
which is a clear experimental advantage.

To study the effect of interlayer coupling on superconducting properties of \agf, we interpolate between the bulk and monolayer limits as follow, 
\begin{equation}
H_0=H^{\mathrm{intra}}+\alpha H^{\mathrm{inter}}
\end{equation}
Here, $H^{\mathrm{intra}}$ is a tight binding model within a single layer of \agf, while $H^{\mathrm{inter}}$ is the interlayer hopping term, which is 
scaled by $\alpha$ $\in$ [0, 1].  
Bulk \agf~can be obtained with $\alpha=1$, and $\alpha=0$ correspond to 
a single layer of \agf~\footnote{
A Hubbard $U=0.37$ eV is used for the monolayer calculation, as the reduced screening leads to divergent susceptibility if the bulk $U$ value is used. The results here only show the qualitative trend.}.

\begin{figure}[ht]
\centering
\includegraphics[width=8.cm]{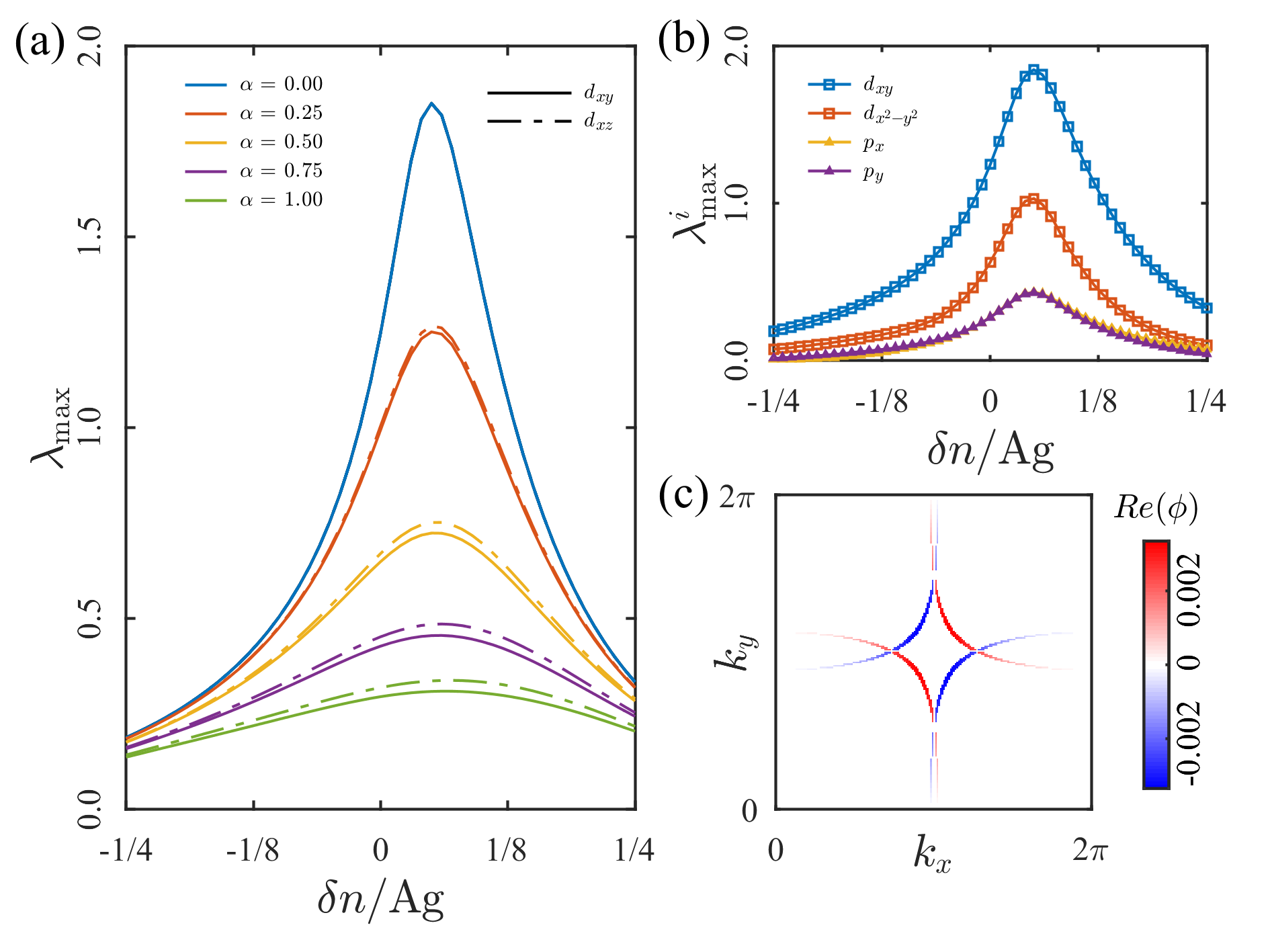}
\caption{(a) Doping dependence of $\lambda^{d_{xy}}_{\mathrm{max}}$ and $\lambda^{d_{xz}}_{\mathrm{max}}$ for different interlayer coupling $\alpha$ at $U=0.37$ eV and $T=30$ meV. 
(b) Doping dependence of $\lambda^i_{\mathrm{max}}$ of various pairing symmetry in single layer \agf. (c) Plot of order parameter $\phi(\bm{k})$  in $k_x$-$k_y$ plane at Fermi surface in 
 single layer \agf~with $\delta n/$Ag$=0.05$. }
\label{fig:4}
\end{figure}

In Fig.~\ref{fig:4}(a), $\lambda^{d_{xy}}_{\mathrm{max}}$ and $\lambda^{d_{xz}}_{\mathrm{max}}$ as functions of doping concentrations for different interlayer coupling strength $\alpha$ are shown. 
It can be seen that the interlayer coupling tends to suppress the superconducting pairing in both hole and electron doping 
because a stronger interlayer coupling amounts to a weaker electronic correlations as discussed earlier. 
Evidently the quasi-2D nature of the crystal structure is one of the crucial factors in the favor of high $T_\text{C}$ superconductivity,
 which again confirms the resemblance to cuprates. 
The optimal electron doping concentration remains unchanged as of bulk \agf, which indicates that the Fermi surface nesting responsible for 
divergence of spin susceptibility mainly occur within a single layer without any significant interlayer contribution.
In the absence of interlayer coupling, the seperated monolayers have the same leading pairing symmetry of the $d_{xy}$-type.
These two degenerate $d_{xy}$ wave then split into $d_{xz}$ and $d_{xy}$ immediately after the interlayer coupling was switched on.
The difference of pairing strength between leading $d_{xz}$ wave and competing $d_{xy}$ was also found to increase with increase of $\alpha$. 

For single layer \agf, the symmetry reduces to $C_{2h}$ point group from $D_{2h}$ of bulk \agf. 
Performing similar analysis to that for bulk, we show the doping dependence of $\lambda^i_{\mathrm{max}}$ for several pairing symmetry 
for single layer in Fig.~\ref{fig:4}(b). As seen previously in Fig.~\ref{fig:4}(a), the leading pairing symmetry is  $d_{xy}$ followed by $d_{x^2-y^2}$ wave. The 
real part of the $\phi(\bm{k})$ corresponding to the $\lambda_{\mathrm{max}}$ at the optimal doping concentration is shown in Fig.~\ref{fig:4}(c),
 which clearly reveals $d_{xy}$ pairing symmetry in the monolayer limit. 

\textsf{\textit{Summary and outlook}}.
Our calculations indicate AgF$_2$ is not only chemically exotic, but also harbors unconventional superconductivity in a way very similar to  high-$T_\text{C}$ cuprates. Our multiband Hubbard model reveals a magnetic instability in accordance with the experimentally obtained magnetic ground state. In the fluctuation-exchange approximation, we find a superconducting ground state with a singlet $d$-wave pairing for the bulk \agf~at  an optimal electron doping of 5\%. By varying the strength of interlayer interaction, we show that the superconducting pairing strength increases with decreasing interlayer coupling, highlighting  the crucial role played by quasi-2D crystal structure on superconducting properties of such materials.

Drawing hints from cuprates, metallization of bulk \agf~can be achieved by synthesizing it with a modified composition as is  done in case of La$_{2-x}$Ba$_x$CuO$_4$/La$_{2-x}$Sr$_x$CuO$_4$ \cite{cu1,cuprate_dope_nat},
leading to doping of extra charge carriers in transition metal-anion plane. Another route to  metallization is electric gating \cite{gating2}. A monolayer of \agf~may be realized by epitaxial growth \cite{epitaxy}, metallization of which can be achieved during the deposition process. The idea of liquid-gating induced superconductivity in thin films \cite{gating_mono} can also be applied to monolayer \agf.

\begin{acknowledgments}
This work is supported by the National Natural Science Foundation of China (Grant No. 11725415 and No. 11934001), the Ministry of Science and Technology of China (Grant No. 2018YFA0305601 and No. 2016YFA0301004), and the Strategic Priority Research Program of the Chinese Academy of Sciences (Grant No. XDB28000000). 
\end{acknowledgments}


\begin{thebibliography}{31}%
\makeatletter
\providecommand \@ifxundefined [1]{%
 \@ifx{#1\undefined}
}%
\providecommand \@ifnum [1]{%
 \ifnum #1\expandafter \@firstoftwo
 \else \expandafter \@secondoftwo
 \fi
}%
\providecommand \@ifx [1]{%
 \ifx #1\expandafter \@firstoftwo
 \else \expandafter \@secondoftwo
 \fi
}%
\providecommand \natexlab [1]{#1}%
\providecommand \enquote  [1]{``#1''}%
\providecommand \bibnamefont  [1]{#1}%
\providecommand \bibfnamefont [1]{#1}%
\providecommand \citenamefont [1]{#1}%
\providecommand \href@noop [0]{\@secondoftwo}%
\providecommand \href [0]{\begingroup \@sanitize@url \@href}%
\providecommand \@href[1]{\@@startlink{#1}\@@href}%
\providecommand \@@href[1]{\endgroup#1\@@endlink}%
\providecommand \@sanitize@url [0]{\catcode `\\12\catcode `\$12\catcode
  `\&12\catcode `\#12\catcode `\^12\catcode `\_12\catcode `\%12\relax}%
\providecommand \@@startlink[1]{}%
\providecommand \@@endlink[0]{}%
\providecommand \url  [0]{\begingroup\@sanitize@url \@url }%
\providecommand \@url [1]{\endgroup\@href {#1}{\urlprefix }}%
\providecommand \urlprefix  [0]{URL }%
\providecommand \Eprint [0]{\href }%
\providecommand \doibase [0]{http://dx.doi.org/}%
\providecommand \selectlanguage [0]{\@gobble}%
\providecommand \bibinfo  [0]{\@secondoftwo}%
\providecommand \bibfield  [0]{\@secondoftwo}%
\providecommand \translation [1]{[#1]}%
\providecommand \BibitemOpen [0]{}%
\providecommand \bibitemStop [0]{}%
\providecommand \bibitemNoStop [0]{.\EOS\space}%
\providecommand \EOS [0]{\spacefactor3000\relax}%
\providecommand \BibitemShut  [1]{\csname bibitem#1\endcsname}%
\let\auto@bib@innerbib\@empty
\bibitem [{\citenamefont {Bednorz}\ and\ \citenamefont {Müller}(1986)}]{cu1}%
  \BibitemOpen
  \bibfield  {author} {\bibinfo {author} {\bibfnamefont {J.~G.}\ \bibnamefont
  {Bednorz}}\ and\ \bibinfo {author} {\bibfnamefont {K.~A.}\ \bibnamefont
  {Müller}},\ }\href {\doibase 10.1007/BF01303701} {\bibfield  {journal}
  {\bibinfo  {journal} {Z. Phy. B - Cond. Mat.}\ }\textbf {\bibinfo {volume}
  {64}},\ \bibinfo {pages} {189} (\bibinfo {year} {1986})}\BibitemShut
  {NoStop}%
\bibitem [{\citenamefont {Bednorz}\ \emph {et~al.}(1987)\citenamefont
  {Bednorz}, \citenamefont {Takashige},\ and\ \citenamefont {Müller}}]{cu2}%
  \BibitemOpen
  \bibfield  {author} {\bibinfo {author} {\bibfnamefont {J.~G.}\ \bibnamefont
  {Bednorz}}, \bibinfo {author} {\bibfnamefont {M.}~\bibnamefont {Takashige}},
  \ and\ \bibinfo {author} {\bibfnamefont {K.~A.}\ \bibnamefont {Müller}},\
  }\href {\doibase 10.1209/0295-5075/3/3/021} {\bibfield  {journal} {\bibinfo
  {journal} {Europhysics Letters ({EPL})}\ }\textbf {\bibinfo {volume} {3}},\
  \bibinfo {pages} {379} (\bibinfo {year} {1987})}\BibitemShut {NoStop}%
\bibitem [{\citenamefont {Bednorz}\ and\ \citenamefont {M\"uller}(1988)}]{cu3}%
  \BibitemOpen
  \bibfield  {author} {\bibinfo {author} {\bibfnamefont {J.~G.}\ \bibnamefont
  {Bednorz}}\ and\ \bibinfo {author} {\bibfnamefont {K.~A.}\ \bibnamefont
  {M\"uller}},\ }\href {\doibase 10.1103/RevModPhys.60.585} {\bibfield
  {journal} {\bibinfo  {journal} {Rev. Mod. Phys.}\ }\textbf {\bibinfo {volume}
  {60}},\ \bibinfo {pages} {585} (\bibinfo {year} {1988})}\BibitemShut
  {NoStop}%
\bibitem [{\citenamefont {Dagotto}(1994)}]{cu4}%
  \BibitemOpen
  \bibfield  {author} {\bibinfo {author} {\bibfnamefont {E.}~\bibnamefont
  {Dagotto}},\ }\href {\doibase 10.1103/RevModPhys.66.763} {\bibfield
  {journal} {\bibinfo  {journal} {Rev. Mod. Phys.}\ }\textbf {\bibinfo {volume}
  {66}},\ \bibinfo {pages} {763} (\bibinfo {year} {1994})}\BibitemShut
  {NoStop}%
\bibitem [{\citenamefont {Kastner}\ \emph {et~al.}(1998)\citenamefont
  {Kastner}, \citenamefont {Birgeneau}, \citenamefont {Shirane},\ and\
  \citenamefont {Endoh}}]{cu5}%
  \BibitemOpen
  \bibfield  {author} {\bibinfo {author} {\bibfnamefont {M.~A.}\ \bibnamefont
  {Kastner}}, \bibinfo {author} {\bibfnamefont {R.~J.}\ \bibnamefont
  {Birgeneau}}, \bibinfo {author} {\bibfnamefont {G.}~\bibnamefont {Shirane}},
  \ and\ \bibinfo {author} {\bibfnamefont {Y.}~\bibnamefont {Endoh}},\ }\href
  {\doibase 10.1103/RevModPhys.70.897} {\bibfield  {journal} {\bibinfo
  {journal} {Rev. Mod. Phys.}\ }\textbf {\bibinfo {volume} {70}},\ \bibinfo
  {pages} {897} (\bibinfo {year} {1998})}\BibitemShut {NoStop}%
\bibitem [{\citenamefont {Scalapino}(2012)}]{cu6}%
  \BibitemOpen
  \bibfield  {author} {\bibinfo {author} {\bibfnamefont {D.~J.}\ \bibnamefont
  {Scalapino}},\ }\href {\doibase 10.1103/RevModPhys.84.1383} {\bibfield
  {journal} {\bibinfo  {journal} {Rev. Mod. Phys.}\ }\textbf {\bibinfo {volume}
  {84}},\ \bibinfo {pages} {1383} (\bibinfo {year} {2012})}\BibitemShut
  {NoStop}%
\bibitem [{\citenamefont {Stewart}(2011)}]{fe11}%
  \BibitemOpen
  \bibfield  {author} {\bibinfo {author} {\bibfnamefont {G.~R.}\ \bibnamefont
  {Stewart}},\ }\href {\doibase 10.1103/RevModPhys.83.1589} {\bibfield
  {journal} {\bibinfo  {journal} {Rev. Mod. Phys.}\ }\textbf {\bibinfo {volume}
  {83}},\ \bibinfo {pages} {1589} (\bibinfo {year} {2011})}\BibitemShut
  {NoStop}%
\bibitem [{\citenamefont {Cao}\ \emph {et~al.}(2018)\citenamefont {Cao},
  \citenamefont {Fatemi}, \citenamefont {Fang}, \citenamefont {Watanabe},
  \citenamefont {Taniguchi}, \citenamefont {Kaxiras},\ and\ \citenamefont
  {Jarillo-Herrero}}]{bg}%
  \BibitemOpen
  \bibfield  {author} {\bibinfo {author} {\bibfnamefont {Y.}~\bibnamefont
  {Cao}}, \bibinfo {author} {\bibfnamefont {V.}~\bibnamefont {Fatemi}},
  \bibinfo {author} {\bibfnamefont {S.}~\bibnamefont {Fang}}, \bibinfo {author}
  {\bibfnamefont {K.}~\bibnamefont {Watanabe}}, \bibinfo {author}
  {\bibfnamefont {T.}~\bibnamefont {Taniguchi}}, \bibinfo {author}
  {\bibfnamefont {E.}~\bibnamefont {Kaxiras}}, \ and\ \bibinfo {author}
  {\bibfnamefont {P.}~\bibnamefont {Jarillo-Herrero}},\ }\href {\doibase
  10.1038/nature26160} {\bibfield  {journal} {\bibinfo  {journal} {Nature}\
  }\textbf {\bibinfo {volume} {556}},\ \bibinfo {pages} {43} (\bibinfo {year}
  {2018})}\BibitemShut {NoStop}%
\bibitem [{\citenamefont {Gawraczy{\'n}ski}\ \emph {et~al.}(2019)\citenamefont
  {Gawraczy{\'n}ski}, \citenamefont {Kurzyd{\l}owski}, \citenamefont {Ewings},
  \citenamefont {Bandaru}, \citenamefont {Gadomski}, \citenamefont {Mazej},
  \citenamefont {Ruani}, \citenamefont {Bergenti}, \citenamefont {Jaro{\'n}},
  \citenamefont {Ozarowski}, \citenamefont {Hill}, \citenamefont
  {Leszczy{\'n}ski}, \citenamefont {Tok{\'a}r}, \citenamefont {Derzsi},
  \citenamefont {Barone}, \citenamefont {Wohlfeld}, \citenamefont {Lorenzana},\
  and\ \citenamefont {Grochala}}]{agf1}%
  \BibitemOpen
  \bibfield  {author} {\bibinfo {author} {\bibfnamefont {J.}~\bibnamefont
  {Gawraczy{\'n}ski}}, \bibinfo {author} {\bibfnamefont {D.}~\bibnamefont
  {Kurzyd{\l}owski}}, \bibinfo {author} {\bibfnamefont {R.~A.}\ \bibnamefont
  {Ewings}}, \bibinfo {author} {\bibfnamefont {S.}~\bibnamefont {Bandaru}},
  \bibinfo {author} {\bibfnamefont {W.}~\bibnamefont {Gadomski}}, \bibinfo
  {author} {\bibfnamefont {Z.}~\bibnamefont {Mazej}}, \bibinfo {author}
  {\bibfnamefont {G.}~\bibnamefont {Ruani}}, \bibinfo {author} {\bibfnamefont
  {I.}~\bibnamefont {Bergenti}}, \bibinfo {author} {\bibfnamefont
  {T.}~\bibnamefont {Jaro{\'n}}}, \bibinfo {author} {\bibfnamefont
  {A.}~\bibnamefont {Ozarowski}}, \bibinfo {author} {\bibfnamefont
  {S.}~\bibnamefont {Hill}}, \bibinfo {author} {\bibfnamefont {P.~J.}\
  \bibnamefont {Leszczy{\'n}ski}}, \bibinfo {author} {\bibfnamefont
  {K.}~\bibnamefont {Tok{\'a}r}}, \bibinfo {author} {\bibfnamefont
  {M.}~\bibnamefont {Derzsi}}, \bibinfo {author} {\bibfnamefont
  {P.}~\bibnamefont {Barone}}, \bibinfo {author} {\bibfnamefont
  {K.}~\bibnamefont {Wohlfeld}}, \bibinfo {author} {\bibfnamefont
  {J.}~\bibnamefont {Lorenzana}}, \ and\ \bibinfo {author} {\bibfnamefont
  {W.}~\bibnamefont {Grochala}},\ }\href {\doibase 10.1073/pnas.1812857116}
  {\bibfield  {journal} {\bibinfo  {journal} {Proc. Natl. Acad. Sci. USA}\
  }\textbf {\bibinfo {volume} {116}},\ \bibinfo {pages} {1495} (\bibinfo {year}
  {2019})}\BibitemShut {NoStop}%
\bibitem [{\citenamefont {Miller}\ and\ \citenamefont {Botana}(2020)}]{agf5}%
  \BibitemOpen
  \bibfield  {author} {\bibinfo {author} {\bibfnamefont {C.}~\bibnamefont
  {Miller}}\ and\ \bibinfo {author} {\bibfnamefont {A.~S.}\ \bibnamefont
  {Botana}},\ }\href {\doibase 10.1103/PhysRevB.101.195116} {\bibfield
  {journal} {\bibinfo  {journal} {Phys. Rev. B}\ }\textbf {\bibinfo {volume}
  {101}},\ \bibinfo {pages} {195116} (\bibinfo {year} {2020})}\BibitemShut
  {NoStop}%
\bibitem [{\citenamefont {Charpin}\ \emph {et~al.}(1970)\citenamefont
  {Charpin}, \citenamefont {Plurien},\ and\ \citenamefont {Meriel}}]{agf2}%
  \BibitemOpen
  \bibfield  {author} {\bibinfo {author} {\bibfnamefont {P.}~\bibnamefont
  {Charpin}}, \bibinfo {author} {\bibfnamefont {P.}~\bibnamefont {Plurien}}, \
  and\ \bibinfo {author} {\bibfnamefont {P.}~\bibnamefont {Meriel}},\
  }\href@noop {} {\bibfield  {journal} {\bibinfo  {journal} {Bull. Soc. Fr.
  Miner. Cristallogr.}\ }\textbf {\bibinfo {volume} {93}},\ \bibinfo {pages}
  {7} (\bibinfo {year} {1970})}\BibitemShut {NoStop}%
\bibitem [{\citenamefont {Fischer}\ \emph
  {et~al.}(1971{\natexlab{a}})\citenamefont {Fischer}, \citenamefont
  {Schwarzenbach},\ and\ \citenamefont {Rietveld}}]{agf3}%
  \BibitemOpen
  \bibfield  {author} {\bibinfo {author} {\bibfnamefont {P.}~\bibnamefont
  {Fischer}}, \bibinfo {author} {\bibfnamefont {D.}~\bibnamefont
  {Schwarzenbach}}, \ and\ \bibinfo {author} {\bibfnamefont {H.}~\bibnamefont
  {Rietveld}},\ }\href {\doibase https://doi.org/10.1016/0022-3697(71)90003-5}
  {\bibfield  {journal} {\bibinfo  {journal} {J. Phys. Chem. Solids}\ }\textbf
  {\bibinfo {volume} {32}},\ \bibinfo {pages} {543} (\bibinfo {year}
  {1971}{\natexlab{a}})}\BibitemShut {NoStop}%
\bibitem [{\citenamefont {Armitage}\ \emph {et~al.}(2010)\citenamefont
  {Armitage}, \citenamefont {Fournier},\ and\ \citenamefont
  {Greene}}]{cu7_dope}%
  \BibitemOpen
  \bibfield  {author} {\bibinfo {author} {\bibfnamefont {N.~P.}\ \bibnamefont
  {Armitage}}, \bibinfo {author} {\bibfnamefont {P.}~\bibnamefont {Fournier}},
  \ and\ \bibinfo {author} {\bibfnamefont {R.~L.}\ \bibnamefont {Greene}},\
  }\href {\doibase 10.1103/RevModPhys.82.2421} {\bibfield  {journal} {\bibinfo
  {journal} {Rev. Mod. Phys.}\ }\textbf {\bibinfo {volume} {82}},\ \bibinfo
  {pages} {2421} (\bibinfo {year} {2010})}\BibitemShut {NoStop}%
\bibitem [{\citenamefont {Fischer}\ \emph
  {et~al.}(1971{\natexlab{b}})\citenamefont {Fischer}, \citenamefont {Roult},\
  and\ \citenamefont {Schwarzenbach}}]{agf4}%
  \BibitemOpen
  \bibfield  {author} {\bibinfo {author} {\bibfnamefont {P.}~\bibnamefont
  {Fischer}}, \bibinfo {author} {\bibfnamefont {G.}~\bibnamefont {Roult}}, \
  and\ \bibinfo {author} {\bibfnamefont {D.}~\bibnamefont {Schwarzenbach}},\
  }\href {\doibase https://doi.org/10.1016/S0022-3697(71)80057-4} {\bibfield
  {journal} {\bibinfo  {journal} {J. Phys. Chem. Solids}\ }\textbf {\bibinfo
  {volume} {32}},\ \bibinfo {pages} {1641 } (\bibinfo {year}
  {1971}{\natexlab{b}})}\BibitemShut {NoStop}%
\bibitem [{sm()}]{sm}%
  \BibitemOpen
  \href@noop {} {}\bibinfo {note} {See Supplemental Material at XXX for details
  about the methodology of $ab$ $initio$, random-phase approximation and
  fluctuation exchange approximation calculations, character table for $D_{2h}$
  and $C_{2h}$ point group symmetry and Fig. S1}\BibitemShut {NoStop}%
\bibitem [{\citenamefont {Mostofi}\ \emph {et~al.}(2008)\citenamefont
  {Mostofi}, \citenamefont {Yates}, \citenamefont {Lee}, \citenamefont {Souza},
  \citenamefont {Vanderbilt},\ and\ \citenamefont {Marzari}}]{wannier90}%
  \BibitemOpen
  \bibfield  {author} {\bibinfo {author} {\bibfnamefont {A.~A.}\ \bibnamefont
  {Mostofi}}, \bibinfo {author} {\bibfnamefont {J.~R.}\ \bibnamefont {Yates}},
  \bibinfo {author} {\bibfnamefont {Y.-S.}\ \bibnamefont {Lee}}, \bibinfo
  {author} {\bibfnamefont {I.}~\bibnamefont {Souza}}, \bibinfo {author}
  {\bibfnamefont {D.}~\bibnamefont {Vanderbilt}}, \ and\ \bibinfo {author}
  {\bibfnamefont {N.}~\bibnamefont {Marzari}},\ }\href {\doibase
  10.1016/j.cpc.2007.11.016} {\bibfield  {journal} {\bibinfo  {journal}
  {Comput. Phys. Commun.}\ }\textbf {\bibinfo {volume} {178}},\ \bibinfo
  {pages} {685 } (\bibinfo {year} {2008})}\BibitemShut {NoStop}%
\bibitem [{\citenamefont {Graser}\ \emph {et~al.}(2009)\citenamefont {Graser},
  \citenamefont {Maier}, \citenamefont {Hirschfeld},\ and\ \citenamefont
  {Scalapino}}]{rpa1}%
  \BibitemOpen
  \bibfield  {author} {\bibinfo {author} {\bibfnamefont {S.}~\bibnamefont
  {Graser}}, \bibinfo {author} {\bibfnamefont {T.~A.}\ \bibnamefont {Maier}},
  \bibinfo {author} {\bibfnamefont {P.~J.}\ \bibnamefont {Hirschfeld}}, \ and\
  \bibinfo {author} {\bibfnamefont {D.~J.}\ \bibnamefont {Scalapino}},\ }\href
  {\doibase 10.1088/1367-2630/11/2/025016} {\bibfield  {journal} {\bibinfo
  {journal} {New J. Phys.}\ }\textbf {\bibinfo {volume} {11}},\ \bibinfo
  {pages} {025016} (\bibinfo {year} {2009})}\BibitemShut {NoStop}%
\bibitem [{\citenamefont {Maier}\ \emph {et~al.}(2011)\citenamefont {Maier},
  \citenamefont {Graser}, \citenamefont {Hirschfeld},\ and\ \citenamefont
  {Scalapino}}]{rpa2}%
  \BibitemOpen
  \bibfield  {author} {\bibinfo {author} {\bibfnamefont {T.~A.}\ \bibnamefont
  {Maier}}, \bibinfo {author} {\bibfnamefont {S.}~\bibnamefont {Graser}},
  \bibinfo {author} {\bibfnamefont {P.~J.}\ \bibnamefont {Hirschfeld}}, \ and\
  \bibinfo {author} {\bibfnamefont {D.~J.}\ \bibnamefont {Scalapino}},\ }\href
  {\doibase 10.1103/PhysRevB.83.100515} {\bibfield  {journal} {\bibinfo
  {journal} {Phys. Rev. B}\ }\textbf {\bibinfo {volume} {83}},\ \bibinfo
  {pages} {100515} (\bibinfo {year} {2011})}\BibitemShut {NoStop}%
\bibitem [{\citenamefont {Kuroki}\ \emph {et~al.}(2008)\citenamefont {Kuroki},
  \citenamefont {Onari}, \citenamefont {Arita}, \citenamefont {Usui},
  \citenamefont {Tanaka}, \citenamefont {Kontani},\ and\ \citenamefont
  {Aoki}}]{rpa3}%
  \BibitemOpen
  \bibfield  {author} {\bibinfo {author} {\bibfnamefont {K.}~\bibnamefont
  {Kuroki}}, \bibinfo {author} {\bibfnamefont {S.}~\bibnamefont {Onari}},
  \bibinfo {author} {\bibfnamefont {R.}~\bibnamefont {Arita}}, \bibinfo
  {author} {\bibfnamefont {H.}~\bibnamefont {Usui}}, \bibinfo {author}
  {\bibfnamefont {Y.}~\bibnamefont {Tanaka}}, \bibinfo {author} {\bibfnamefont
  {H.}~\bibnamefont {Kontani}}, \ and\ \bibinfo {author} {\bibfnamefont
  {H.}~\bibnamefont {Aoki}},\ }\href {\doibase 10.1103/PhysRevLett.101.087004}
  {\bibfield  {journal} {\bibinfo  {journal} {Phys. Rev. Lett.}\ }\textbf
  {\bibinfo {volume} {101}},\ \bibinfo {pages} {087004} (\bibinfo {year}
  {2008})}\BibitemShut {NoStop}%
\bibitem [{\citenamefont {Wu}\ \emph {et~al.}(2015)\citenamefont {Wu},
  \citenamefont {Yang}, \citenamefont {Le}, \citenamefont {Fan},\ and\
  \citenamefont {Hu}}]{rpa4}%
  \BibitemOpen
  \bibfield  {author} {\bibinfo {author} {\bibfnamefont {X.}~\bibnamefont
  {Wu}}, \bibinfo {author} {\bibfnamefont {F.}~\bibnamefont {Yang}}, \bibinfo
  {author} {\bibfnamefont {C.}~\bibnamefont {Le}}, \bibinfo {author}
  {\bibfnamefont {H.}~\bibnamefont {Fan}}, \ and\ \bibinfo {author}
  {\bibfnamefont {J.}~\bibnamefont {Hu}},\ }\href {\doibase
  10.1103/PhysRevB.92.104511} {\bibfield  {journal} {\bibinfo  {journal} {Phys.
  Rev. B}\ }\textbf {\bibinfo {volume} {92}},\ \bibinfo {pages} {104511}
  (\bibinfo {year} {2015})}\BibitemShut {NoStop}%
\bibitem [{\citenamefont {Kuroki}\ \emph {et~al.}(2009)\citenamefont {Kuroki},
  \citenamefont {Usui}, \citenamefont {Onari}, \citenamefont {Arita},\ and\
  \citenamefont {Aoki}}]{rpa5}%
  \BibitemOpen
  \bibfield  {author} {\bibinfo {author} {\bibfnamefont {K.}~\bibnamefont
  {Kuroki}}, \bibinfo {author} {\bibfnamefont {H.}~\bibnamefont {Usui}},
  \bibinfo {author} {\bibfnamefont {S.}~\bibnamefont {Onari}}, \bibinfo
  {author} {\bibfnamefont {R.}~\bibnamefont {Arita}}, \ and\ \bibinfo {author}
  {\bibfnamefont {H.}~\bibnamefont {Aoki}},\ }\href {\doibase
  10.1103/PhysRevB.79.224511} {\bibfield  {journal} {\bibinfo  {journal} {Phys.
  Rev. B}\ }\textbf {\bibinfo {volume} {79}},\ \bibinfo {pages} {224511}
  (\bibinfo {year} {2009})}\BibitemShut {NoStop}%
\bibitem [{\citenamefont {Bickers}\ \emph {et~al.}(1989)\citenamefont
  {Bickers}, \citenamefont {Scalapino},\ and\ \citenamefont {White}}]{FLEX1}%
  \BibitemOpen
  \bibfield  {author} {\bibinfo {author} {\bibfnamefont {N.~E.}\ \bibnamefont
  {Bickers}}, \bibinfo {author} {\bibfnamefont {D.~J.}\ \bibnamefont
  {Scalapino}}, \ and\ \bibinfo {author} {\bibfnamefont {S.~R.}\ \bibnamefont
  {White}},\ }\href {\doibase 10.1103/PhysRevLett.62.961} {\bibfield  {journal}
  {\bibinfo  {journal} {Phys. Rev. Lett.}\ }\textbf {\bibinfo {volume} {62}},\
  \bibinfo {pages} {961} (\bibinfo {year} {1989})}\BibitemShut {NoStop}%
\bibitem [{\citenamefont {Takimoto}\ \emph {et~al.}(2004)\citenamefont
  {Takimoto}, \citenamefont {Hotta},\ and\ \citenamefont {Ueda}}]{FLEX2}%
  \BibitemOpen
  \bibfield  {author} {\bibinfo {author} {\bibfnamefont {T.}~\bibnamefont
  {Takimoto}}, \bibinfo {author} {\bibfnamefont {T.}~\bibnamefont {Hotta}}, \
  and\ \bibinfo {author} {\bibfnamefont {K.}~\bibnamefont {Ueda}},\ }\href
  {\doibase 10.1103/PhysRevB.69.104504} {\bibfield  {journal} {\bibinfo
  {journal} {Phys. Rev. B}\ }\textbf {\bibinfo {volume} {69}},\ \bibinfo
  {pages} {104504} (\bibinfo {year} {2004})}\BibitemShut {NoStop}%
\bibitem [{\citenamefont {Jiang}\ \emph {et~al.}(2014)\citenamefont {Jiang},
  \citenamefont {Hu}, \citenamefont {You}, \citenamefont {Li}, \citenamefont
  {Li}, \citenamefont {Wang}, \citenamefont {Mu}, \citenamefont {Chen},
  \citenamefont {Zhang}, \citenamefont {Yu}, \citenamefont {Zhu}, \citenamefont
  {Sun}, \citenamefont {Lin}, \citenamefont {Xiao}, \citenamefont {Xie},\ and\
  \citenamefont {Jiang}}]{cu_mono}%
  \BibitemOpen
  \bibfield  {author} {\bibinfo {author} {\bibfnamefont {D.}~\bibnamefont
  {Jiang}}, \bibinfo {author} {\bibfnamefont {T.}~\bibnamefont {Hu}}, \bibinfo
  {author} {\bibfnamefont {L.}~\bibnamefont {You}}, \bibinfo {author}
  {\bibfnamefont {Q.}~\bibnamefont {Li}}, \bibinfo {author} {\bibfnamefont
  {A.}~\bibnamefont {Li}}, \bibinfo {author} {\bibfnamefont {H.}~\bibnamefont
  {Wang}}, \bibinfo {author} {\bibfnamefont {G.}~\bibnamefont {Mu}}, \bibinfo
  {author} {\bibfnamefont {Z.}~\bibnamefont {Chen}}, \bibinfo {author}
  {\bibfnamefont {H.}~\bibnamefont {Zhang}}, \bibinfo {author} {\bibfnamefont
  {G.}~\bibnamefont {Yu}}, \bibinfo {author} {\bibfnamefont {J.}~\bibnamefont
  {Zhu}}, \bibinfo {author} {\bibfnamefont {Q.}~\bibnamefont {Sun}}, \bibinfo
  {author} {\bibfnamefont {C.}~\bibnamefont {Lin}}, \bibinfo {author}
  {\bibfnamefont {H.}~\bibnamefont {Xiao}}, \bibinfo {author} {\bibfnamefont
  {X.}~\bibnamefont {Xie}}, \ and\ \bibinfo {author} {\bibfnamefont
  {M.}~\bibnamefont {Jiang}},\ }\href {\doibase 10.1038/ncomms6708} {\bibfield
  {journal} {\bibinfo  {journal} {Nat. Commun.}\ }\textbf {\bibinfo {volume}
  {5}},\ \bibinfo {pages} {5708} (\bibinfo {year} {2014})}\BibitemShut
  {NoStop}%
\bibitem [{\citenamefont {Yu}\ \emph {et~al.}(2019)\citenamefont {Yu},
  \citenamefont {Ma}, \citenamefont {Cai}, \citenamefont {Zhong}, \citenamefont
  {Ye}, \citenamefont {Shen}, \citenamefont {Gu}, \citenamefont {Chen},\ and\
  \citenamefont {Zhang}}]{mono}%
  \BibitemOpen
  \bibfield  {author} {\bibinfo {author} {\bibfnamefont {Y.}~\bibnamefont
  {Yu}}, \bibinfo {author} {\bibfnamefont {L.}~\bibnamefont {Ma}}, \bibinfo
  {author} {\bibfnamefont {P.}~\bibnamefont {Cai}}, \bibinfo {author}
  {\bibfnamefont {R.}~\bibnamefont {Zhong}}, \bibinfo {author} {\bibfnamefont
  {C.}~\bibnamefont {Ye}}, \bibinfo {author} {\bibfnamefont {J.}~\bibnamefont
  {Shen}}, \bibinfo {author} {\bibfnamefont {G.~D.}\ \bibnamefont {Gu}},
  \bibinfo {author} {\bibfnamefont {X.~H.}\ \bibnamefont {Chen}}, \ and\
  \bibinfo {author} {\bibfnamefont {Y.}~\bibnamefont {Zhang}},\ }\href
  {\doibase 10.1038/s41586-019-1718-x} {\bibfield  {journal} {\bibinfo
  {journal} {Nature}\ }\textbf {\bibinfo {volume} {575}},\ \bibinfo {pages}
  {156} (\bibinfo {year} {2019})}\BibitemShut {NoStop}%
\bibitem [{\citenamefont {Ye}\ \emph {et~al.}(2012)\citenamefont {Ye},
  \citenamefont {Zhang}, \citenamefont {Akashi}, \citenamefont {Bahramy},
  \citenamefont {Arita},\ and\ \citenamefont {Iwasa}}]{gating}%
  \BibitemOpen
  \bibfield  {author} {\bibinfo {author} {\bibfnamefont {J.~T.}\ \bibnamefont
  {Ye}}, \bibinfo {author} {\bibfnamefont {Y.~J.}\ \bibnamefont {Zhang}},
  \bibinfo {author} {\bibfnamefont {R.}~\bibnamefont {Akashi}}, \bibinfo
  {author} {\bibfnamefont {M.~S.}\ \bibnamefont {Bahramy}}, \bibinfo {author}
  {\bibfnamefont {R.}~\bibnamefont {Arita}}, \ and\ \bibinfo {author}
  {\bibfnamefont {Y.}~\bibnamefont {Iwasa}},\ }\href {\doibase
  10.1126/science.1228006} {\bibfield  {journal} {\bibinfo  {journal}
  {Science}\ }\textbf {\bibinfo {volume} {338}},\ \bibinfo {pages} {1193}
  (\bibinfo {year} {2012})}\BibitemShut {NoStop}%
\bibitem [{Notei()}]{Notei}%
  \BibitemOpen
  \bibinfo {note} {A Hubbard $U=0.37$ eV is used for the monolayer calculation,
  as the reduced screening leads to divergent susceptibility if the bulk $U$
  value is used. The results here only show the qualitative trend.}\BibitemShut
  {Stop}%
\bibitem [{\citenamefont {Bollinger}\ \emph {et~al.}(2011)\citenamefont
  {Bollinger}, \citenamefont {Dubuis}, \citenamefont {Yoon}, \citenamefont
  {Pavuna}, \citenamefont {Misewich},\ and\ \citenamefont
  {Bo{\v{z}}ovi{\'{c}}}}]{cuprate_dope_nat}%
  \BibitemOpen
  \bibfield  {author} {\bibinfo {author} {\bibfnamefont {A.~T.}\ \bibnamefont
  {Bollinger}}, \bibinfo {author} {\bibfnamefont {G.}~\bibnamefont {Dubuis}},
  \bibinfo {author} {\bibfnamefont {J.}~\bibnamefont {Yoon}}, \bibinfo {author}
  {\bibfnamefont {D.}~\bibnamefont {Pavuna}}, \bibinfo {author} {\bibfnamefont
  {J.}~\bibnamefont {Misewich}}, \ and\ \bibinfo {author} {\bibfnamefont
  {I.}~\bibnamefont {Bo{\v{z}}ovi{\'{c}}}},\ }\href {\doibase
  10.1038/nature09998} {\bibfield  {journal} {\bibinfo  {journal} {Nature}\
  }\textbf {\bibinfo {volume} {472}},\ \bibinfo {pages} {458} (\bibinfo {year}
  {2011})}\BibitemShut {NoStop}%
\bibitem [{\citenamefont {Goldman}(2014)}]{gating2}%
  \BibitemOpen
  \bibfield  {author} {\bibinfo {author} {\bibfnamefont {A.}~\bibnamefont
  {Goldman}},\ }\href {\doibase 10.1146/annurev-matsci-070813-113407}
  {\bibfield  {journal} {\bibinfo  {journal} {Annu. Rev. Mater. Res.}\ }\textbf
  {\bibinfo {volume} {44}},\ \bibinfo {pages} {45} (\bibinfo {year}
  {2014})}\BibitemShut {NoStop}%
\bibitem [{\citenamefont {Grzelak}\ \emph {et~al.}(2020)\citenamefont
  {Grzelak}, \citenamefont {Su}, \citenamefont {Yang}, \citenamefont
  {Kurzyd\l{}owski}, \citenamefont {Lorenzana},\ and\ \citenamefont
  {Grochala}}]{epitaxy}%
  \BibitemOpen
  \bibfield  {author} {\bibinfo {author} {\bibfnamefont {A.}~\bibnamefont
  {Grzelak}}, \bibinfo {author} {\bibfnamefont {H.}~\bibnamefont {Su}},
  \bibinfo {author} {\bibfnamefont {X.}~\bibnamefont {Yang}}, \bibinfo {author}
  {\bibfnamefont {D.}~\bibnamefont {Kurzyd\l{}owski}}, \bibinfo {author}
  {\bibfnamefont {J.}~\bibnamefont {Lorenzana}}, \ and\ \bibinfo {author}
  {\bibfnamefont {W.}~\bibnamefont {Grochala}},\ }\href {\doibase
  10.1103/PhysRevMaterials.4.084405} {\bibfield  {journal} {\bibinfo  {journal}
  {Phys. Rev. Materials}\ }\textbf {\bibinfo {volume} {4}},\ \bibinfo {pages}
  {084405} (\bibinfo {year} {2020})}\BibitemShut {NoStop}%
\bibitem [{\citenamefont {Ye}\ \emph {et~al.}(2010)\citenamefont {Ye},
  \citenamefont {Inoue}, \citenamefont {Kobayashi}, \citenamefont {Kasahara},
  \citenamefont {Yuan}, \citenamefont {Shimotani},\ and\ \citenamefont
  {Iwasa}}]{gating_mono}%
  \BibitemOpen
  \bibfield  {author} {\bibinfo {author} {\bibfnamefont {J.~T.}\ \bibnamefont
  {Ye}}, \bibinfo {author} {\bibfnamefont {S.}~\bibnamefont {Inoue}}, \bibinfo
  {author} {\bibfnamefont {K.}~\bibnamefont {Kobayashi}}, \bibinfo {author}
  {\bibfnamefont {Y.}~\bibnamefont {Kasahara}}, \bibinfo {author}
  {\bibfnamefont {H.~T.}\ \bibnamefont {Yuan}}, \bibinfo {author}
  {\bibfnamefont {H.}~\bibnamefont {Shimotani}}, \ and\ \bibinfo {author}
  {\bibfnamefont {Y.}~\bibnamefont {Iwasa}},\ }\href {\doibase
  10.1038/nmat2587} {\bibfield  {journal} {\bibinfo  {journal} {Nature
  Materials}\ }\textbf {\bibinfo {volume} {9}},\ \bibinfo {pages} {125}
  (\bibinfo {year} {2010})}\BibitemShut {NoStop}%
\end{thebibliography}

%

\end{document}


\title{\textit{Supplemental Material}:\\A silver(II) route to  unconventional superconductivity}

\author{Xiaoqiang Liu}
\thanks{Equal contribution}
\affiliation{International Center for Quantum Materials, School of Physics, Peking University, Beijing 100871, China}

\author{Shishir K. Pandey}
\thanks{Equal contribution}
\affiliation{International Center for Quantum Materials, School of Physics, Peking University, Beijing 100871, China}

\author{Ji Feng}\email{jfeng11@pku.edu.cn}
\affiliation{International Center for Quantum Materials, School of Physics, Peking University, Beijing 100871, China}
\affiliation{Collaborative Innovation Center of Quantum Matter, Beijing 100871,
China}
\affiliation{CAS Center for Excellence in Topological Quantum Computation, University of Chinese Academy of Sciences, Beijing 100190, China}

\maketitle
\subsection{A.~~Density-functional theory calculations}
To calculate electronic structure of AgF$_2$, we have performed density-functional theory based 
calculations using Vienna $ab$ $initio$ simulation package(VASP) \cite{Kresse}. The projector-augmented wave potentials are used in our calculations \cite{PAW,PAWpotentials1}, with a $6\times6\times8$ Gamma-centered mesh of k-points and a plane wave cutoff of 800 eV. The results are found to be well converged with respect to k-point mesh and plane wave numbers. The Perdew-Burke-Ernzerhof functional \cite{PBE} is used for the
exchange-correlation functional, and additional electron-electron correlation is included statically by a $U=8$ eV within the GGA+U formalism \cite{dudarev}.
The total energy is calculated self-consistently till the energy difference between successive steps is below 10$^{-5}$ eV.  
The experimental lattice parameters of orthorhombic AgF$_2$ with space group $Pbca$ (No. 61) are |{\bf a}| = 5.101 \AA{}, |{\bf b}| =  5.568 \AA{} and |{\bf c}| = 5.831 \AA{} \cite{agf2}.

Considering the experimentally observed magnetic ground state and  $U$ = 8 eV on Ag-$d$ states, we perform full optimization of lattice parameters. The value of $U$ on Ag-$d$ orbital is chosen to match the experimentally observed Ag magnetic moments of 0.7 $\mu_B$/Ag.
Change in {\bf a} and {\bf b} lattice 
constants were found to be less than 1 \% while the {\bf c}  vector was enhanced by $\sim$1.7\%. Since these values are very close to the experimental ones, we use the experimental 
structure for further study.
Non-spin polarized $ab$ $initio$ band structure was projected onto a tight-binding model using the maximally-localized Wannier functions for the radial part of the wavefunctions using WANNIER90-VASP interface, implemented within VASP \cite{wannier90}.
Mapping is done for the undoped case considering Ag-$d_{x^2-y^2}$ orbitals and the hopping amplitudes are listed in Table \ref{table:hopping}. 

\begin{table}[htbp]
\caption{List of all hopping amplitudes larger than 5 meV in our Wannier function based tight-binding model of bulk AgF$_2$. $\bm{R}$ and $l/l'$ are the translation vector and orbital index respectively. All other hoppings not listed here can be obtained by applying symmetry operations of $Pbca$, the space group of bulk AgF$_2$.}
\label{table:hopping}
\centering
\begin{tabular}{cccc}
\hline
\hline
$\bm{R}$ & \qquad $l$  & \qquad $l'$ & \qquad$t~(meV)$\\
\hline
   (0 ,  0 ,  0)  &  \qquad 1  & \qquad 2  & \qquad -173.1 \\
   (1 , -1 ,  0)  &  \qquad 1  & \qquad 1  & \qquad  34.4 \\
   (0 ,  1 ,  0)  &  \qquad 1  & \qquad 1  & \qquad  34.0 \\
   (0 ,  1 , -1)  &  \qquad 1  & \qquad 1  & \qquad  28.0 \\
   (0 , -1 ,  0)  &  \qquad 1  & \qquad 3  & \qquad  22.8 \\
   (0 ,  0 , -1)  &  \qquad 1  & \qquad 2  & \qquad  18.2 \\
   (1 ,  0 , -1)  &  \qquad 1  & \qquad 1  & \qquad  18.1 \\
   (1 ,  1 ,  0)  &  \qquad 1  & \qquad 1  & \qquad  15.1 \\
   (0 ,  1 ,  0)  &  \qquad 1  & \qquad 3  & \qquad -11.4 \\
   (0 ,  0 ,  0)  &  \qquad 1  & \qquad 4  & \qquad  7.8 \\
   (1 ,  0 ,  1)  &  \qquad 1  & \qquad 1  & \qquad  7.1 \\
   (0 ,  0 ,  0)  &  \qquad 1  & \qquad 3  & \qquad -7.0 \\
   (1 ,  0 ,  0)  &  \qquad 1  & \qquad 4  & \qquad -6.1 \\
   (1 , -1 , -1)  &  \qquad 1  & \qquad 1  & \qquad -5.6 \\
   (1 ,  0 ,  0)  &  \qquad 1  & \qquad 3  & \qquad  5.3 \\
   (1 ,  0 ,  0)  &  \qquad 1  & \qquad 1  & \qquad  5.1 \\
\hline
\hline
\end{tabular}
\end{table}

\subsection{B.~~Random-phase approximation and fluctuation exchange approximation}
We start by setting up the multiband Hubbard model, by adding the Hubbard $U$ term to a tight-binding model described in section A,
\begin{equation}
H=H^{\mathrm{TB}}+H^{\mathrm{hub}}=\sum_{ijll'\sigma}t_{ij}^{ll'}c_{il\sigma}^{\dagger}c_{jl'\sigma}+U\sum_{il}n_{il\uparrow}n_{il\downarrow}
\end{equation}
Here, $c_{il\sigma}$ is annihilation operator of $d_{x^2-y^2}$ orbital on Ag site $l$ in $i^{th}$ unit cell with spin $\sigma$ and $t_{ij}^{ll'}$ is hopping amplitude. $U$ is the on-site Hubbard parameter.

The random-phase approximation(RPA) \cite{rpa1,rpa2,rpa3,rpa4,rpa5} is employed to study the magnetic instability of the system. The RPA spin and charge  susceptibilities are given by 
\begin{equation}
\begin{array}{cc}
\chi^{c}(q)= & \left[1+\chi^{0}(q)U^{c}\right]^{-1}\chi^{0}(q)\\
\chi^{s}(q)= & \left[1-\chi^{0}(q)U^{s}\right]^{-1}\chi^{0}(q)
\end{array}
\end{equation}
respectively, where 
\begin{equation}
\left(U^{c/s}\right)_{l_{3}l_{4}}^{l_{1}l_{2}}=\begin{cases}
U & l_{1}=l_{2}=l_{3}=l_{4}\\
0 & \mathrm{otherwise}
\end{cases}.
\end{equation}
Here $q=(\bm{q},\omega)$.

In above expression the bare susceptibility $\chi^0(q)$ is given by, 
\begin{equation}
\left(\chi^{0}\right)_{l_{3}l_{4}}^{l_{1}l_{2}}(\bm{q},\omega)=-\frac{1}{N}\sum_{\bm{k},\mu\nu}\frac{a_{\nu}^{l_{2}}(\bm{k}+\bm{q})a_{\nu}^{l_{3}*}(\bm{k}+\bm{q})a_{\mu}^{l_{4}}(\bm{k})a_{\mu}^{l_{1}*}(\bm{k})}{\omega+\varepsilon_{\mu\bm{k}}-\varepsilon_{\nu\bm{k}+\bm{q}}+i0^{+}}[f(\varepsilon_{\mu\bm{k}}-\varepsilon_{F})-f(\varepsilon_{\nu\bm{k}+\bm{q}}-\varepsilon_{F})],
\end{equation}
where $N$ is the number of $\bm{k}$-points and $\mu, \nu$ are the band indices. 
$a_{\mu}^l({\bm{k}})$ is the $l$ component of wave function of band $\mu$, obtained from diagonalization of $H^{\mathrm{TB}}$ and $\varepsilon_{\mu}(\bm{k})$ is the corresponding eigenvalues of band $\mu$ at $\bm{k}$-point. $f(E)$ is the Fermi-Dirac distribution and $\varepsilon_{F}$ is Fermi energy.

Under fluctuation exchange approximation(FLEX) \cite{FLEX1,FLEX2}, the pairing vertex describing the effective electron-electron interaction is given by,
\begin{equation}
\Gamma(q)=\gamma U^{s}\chi^{s}(q)U^{s}-\frac{1}{2}U^{c}\chi^{c}(q)U^{c}+\frac{1}{2}(U^{s}+U^{c}),
\end{equation}
with $\gamma=\frac{3}{2}$ for the singlet channel and $\gamma=-\frac{1}{2}$ for the triplet channel.
The singlet vertex is symmetrized as, 
\begin{equation}
\left(\Gamma^{s}\right)_{l_{3}l_{4}}^{l_{1}l_{2}}(k,k')=\frac{1}{2}\left(\left(\Gamma^{s}\right)_{l_{3}l_{4}}^{l_{1}l_{2}}(k,k')+\left(\Gamma^{s}\right)_{l_{3}l_{2}}^{l_{1}l_{4}}(k,-k')\right),
\end{equation}
and the triplet vertex is anti-symmetrized as, 
\begin{equation}
\left(\Gamma^{t}\right)_{l_{3}l_{4}}^{l_{1}l_{2}}(k,k')=\frac{1}{2}\left(\left(\Gamma^{t}\right)_{l_{3}l_{4}}^{l_{1}l_{2}}(k,k')-\left(\Gamma^{t}\right)_{l_{3}l_{2}}^{l_{1}l_{4}}(k,-k')\right),
\end{equation}
Here, $\Gamma(k,k')=\Gamma(q)$ with $q=k-k'$.

The linearized Eliashberg equation will then be solved to obtain the order parameter and superconducting transition temperature, 
\begin{equation}
\lambda\phi_{l_{1}l_{3}}(k)=-\frac{T}{N}\sum_{q}\sum_{l_{2}l_{4}l_{5}l_{6}}\Gamma_{l_{3}l_{4}}^{l_{1}l_{2}}(q)\phi_{l_{5}l_{6}}(k-q)G_{l_{2}l_{5}}(k-q)G_{l_{4}l_{6}}(q-k).
\end{equation}
Here, $\lambda$ is the eigenvalue indicating the pairing strength. The eigenvector $\phi_{l_{1}l_{2}}(k)$ is the order parameter written in term of orbital index $l_1$ and $l_2$. $G_{l_{1}l_{2}}(k)$ is the Matsubara Green function.

In the weak-coupling regime, the pairing vertex is approximated to be frequency-independent i.e. $\Gamma(q)=\Gamma(\bm{q},\omega=0)$.
After summing over the Matsubara frequencies, we obtain the following equation in band basis, 
\begin{equation}
\lambda\phi_{mn}(\bm{k})=-\frac{1}{N}\sum_{\bm{k}'}\sum_{\mu\nu}\left(\Gamma^{\eta}\right)_{\mu\nu}^{mn}(\bm{k},\bm{k}')F_{\mu\nu}(\bm{k}')\phi_{\mu\nu}(\bm{k}'),
\end{equation}
where
\begin{equation}
F_{\mu\nu}(\bm{k})=-\frac{f(\varepsilon_{\mu\bm{k}}-\varepsilon_{F})+f(\varepsilon_{\nu-\bm{k}}-\varepsilon_{F})-1}{\varepsilon_{\mu\bm{k}}+\varepsilon_{\nu-\bm{k}}-2\varepsilon_{F}}.
\end{equation}
The transformation between orbital basis and band basis for the order parameter is
\begin{equation}
\phi_{mn}(\bm{k})=\sum_{l_{1}l_{2}}\phi_{l_{1}l_{2}}(\bm{k})a_{m}^{l_{1}*}(\bm{k})a_{n}^{l_{2}*}(-\bm{k}),
\label{gap}
\end{equation}
and for the vertex,
\begin{equation}
\left(\Gamma^{\eta}\right)_{\mu\nu}^{mn}(\bm{k},\bm{k}')=\sum_{l_{1}l_{2}l_{3}l_{4}}a_{m}^{l_{1}*}(\bm{k})a_{n}^{l_{3}*}(-\bm{k})\left(\Gamma^{\eta}\right)_{l_{3}l_{4}}^{l_{1}l_{2}}(\bm{k}-\bm{k}')a_{\mu}^{l_{2}}(\bm{k}')a_{\nu}^{l_{4}}(-\bm{k}').
\end{equation}

\subsection{C.~~Pairing symmetry}
One can see that $\phi_{mn}(\bm{k})$ is not gauge invariant because of the eigenvector $a_{\mu}^l({\bf k})$ involved in Eq.(\ref{gap}). However, if the following gauge is chosen,
\begin{equation}
a_{n}^{l}(-\bm{k})=a_{n}^{l}(\bm{k})^*
\end{equation}
for the non-magnetic state \cite{gauge}, then the trace of $\phi_{mn}(\bm{k})$
\begin{equation}
\phi(\bm{k})=\sum_n \phi_{nn}(\bm{k})
\end{equation}
or trace taken on Fermi surface
\begin{equation}
\phi(\bm{k})=\sum_n\phi_{nn}(\bm{k})\delta(\varepsilon_{n\bm{k}}-\varepsilon_{\mathrm{F}})
\end{equation}
is gauge invariant.

For the symmetry group of normal state Hamiltonian $H$, which are $D_{2h}$ for bulk AgF$_2$ and $C_{2h}$ for single layer AgF$_2$, $\phi(\bm{k})$ can be identified as one of the irreducible representations of the symmetry group. All the irreducible representation and basis functions of $D_{2h}$ are listed in Table \ref{table:D2h} and $C_{2h}$ in 
Table~\ref{table:C2h}. Note that the $C_{2}$ rotation in $C_{2h}$ is along $x$-axis.

\begin{table}[htbp]
\caption{Character table for $D_{2h}$ point group}
\label{table:D2h}
\centering
\begin{tabular}{|c|c|c|c|c|c|c|c|c|c|c|}
\hline
& \multirow{2}{*}{$E$}  & \multirow{2}{*}{$C_2(z)$} & \multirow{2}{*}{$C_2(y)$} & \multirow{2}{*}{$C_2(x)$} & \multirow{2}{*}{$i$} & \multirow{2}{*}{$\sigma(xy)$} & \multirow{2}{*}{$\sigma(xz)$} & \multirow{2}{*}{$\sigma(yz)$} & Linear, & \multirow{2}{*}{Quadratic}\\
&&&&&&&&& Rotations&\\
\hline
$A_{g} $ & $+1$ & $+1$ & $+1$ & $+1$ & $+1$ & $+1$ & $+1$ & $+1$ & ---   &  $x^2-y^2,z^2 $ \\
\hline
$B_{1g}$ & $+1$ & $+1$ & $-1$ & $-1$ & $+1$ & $+1$ & $-1$ & $-1$ & $R_z$ &  $xy$ \\
\hline
$B_{2g}$ & $+1$ & $-1$ & $+1$ & $-1$ & $+1$ & $-1$ & $+1$ & $-1$ & $R_y$ &  $xz$ \\
\hline
$B_{3g}$ & $+1$ & $-1$ & $-1$ & $+1$ & $+1$ & $-1$ & $-1$ & $+1$ & $R_x$ &  $yz$ \\
\hline
$A_{u} $ & $+1$ & $+1$ & $+1$ & $+1$ & $-1$ & $-1$ & $-1$ & $-1$ & ---   &  --- \\
\hline
$B_{1u}$ & $+1$ & $+1$ & $-1$ & $-1$ & $-1$ & $-1$ & $+1$ & $+1$ & $z$   & --- \\
\hline
$B_{2u}$ & $+1$ & $-1$ & $+1$ & $-1$ & $-1$ & $+1$ & $-1$ & $+1$ & $y$   & --- \\
\hline
$B_{3u}$ & $+1$ & $-1$ & $-1$ & $+1$ & $-1$ & $+1$ & $+1$ & $-1$ & $x$   & --- \\
\hline
\end{tabular}
\end{table}

\begin{table}[htbp]
\caption{Character table for $C_{2h}$ point group}
\label{table:C2h}
\centering
\begin{tabular}{|c|c|c|c|c|c|c|}
\hline
& \multirow{2}{*}{$E$}  & \multirow{2}{*}{$C_2(x)$} & \multirow{2}{*}{$i$} & \multirow{2}{*}{$\sigma_h$} & Linear, & \multirow{2}{*}{Quadratic}\\
&&&&& Rotations&\\
\hline
$A_{g} $ & $+1$ & $+1$ & $+1$ & $+1$ & $R_x$ 	  &  $x^2-y^2,z^2,yz$ \\
\hline
$B_{1g}$ & $+1$ & $-1$ & $+1$ & $-1$ & $R_y,R_z$  &  $xy$,$xz$ \\
\hline
$A_{u} $ & $+1$ & $+1$ & $-1$ & $-1$ & $x$  	  &  --- \\
\hline
$B_{1u}$ & $+1$ & $-1$ & $-1$ & $+1$ & $y,z$   	  & --- \\
\hline
\end{tabular}
\end{table}

\begin{figure}[ht]
\centering
\includegraphics[width=8.cm]{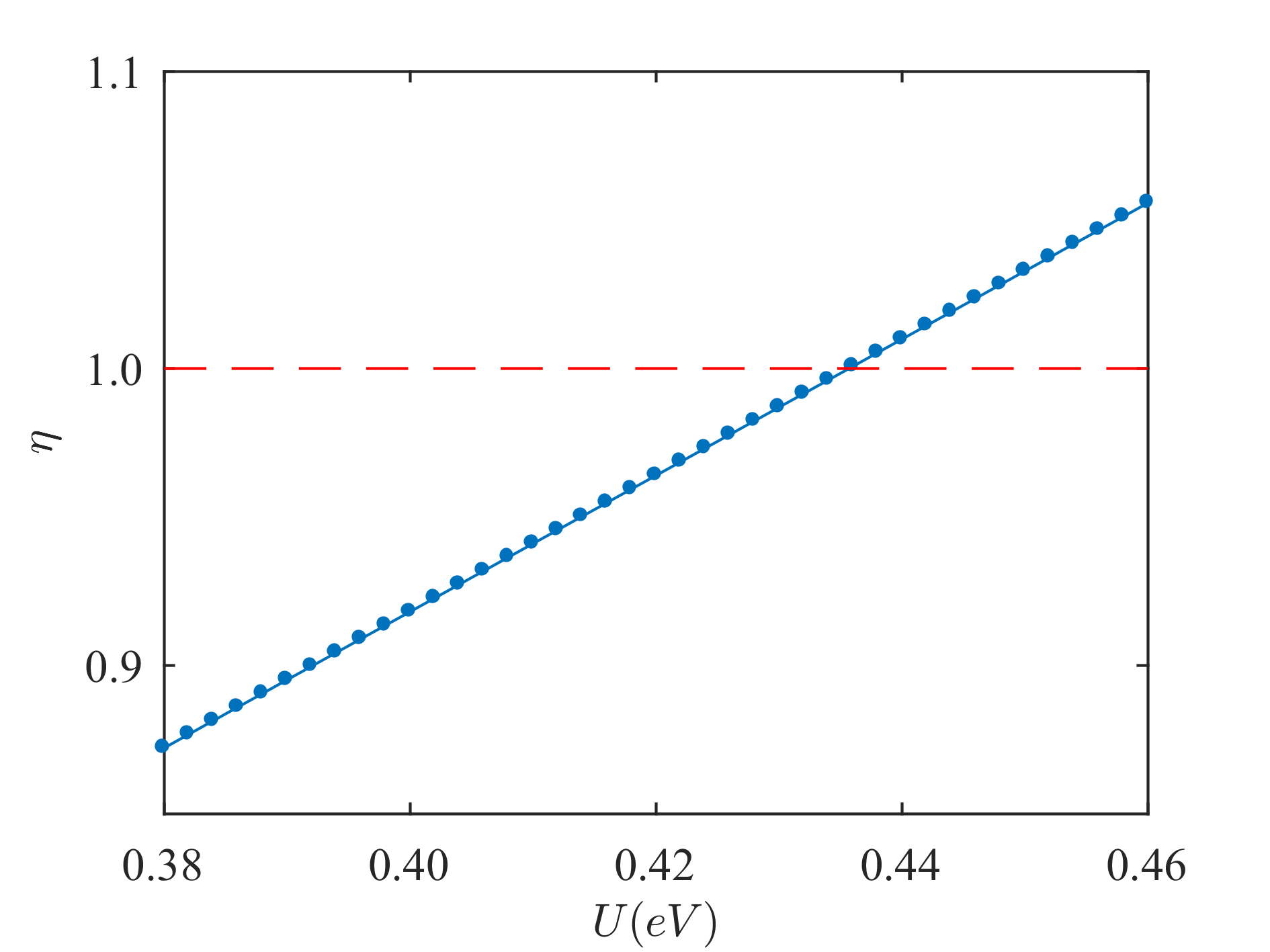}
\caption{The $U$ dependence of largest $\eta$ at experimental N\'eel temperature($\sim$163 K=14 meV). $\eta \approx 1$ when $U=0.44$ eV.}
\label{fig:s1}
\end{figure}

\clearpage

%